\input{aipcheck}

\documentclass[
    ,final            
  ]
  {aipproc}
  
\usepackage{epsfig,graphicx,bbm,psfrag,amssymb}
\usepackage{fancyhdr}

\layoutstyle{6x9}

\begin{document}

\title{Sensitivity to the Higgs sector of SUSY-seesaw models via LFV tau decays}

\classification{{11.30.Hv}, {12.60.Jv}, {14.60.St} , {{\bf Prepint
numbers:} FTUAM-09/19, IFT-UAM/CSIC-09-40}}

\author{M. Herrero}{
  address={Departamento de F\'{\i }sica Te\'{o}rica 
and Instituto de F\'{\i }sica Te\'{o}rica, IFT-UAM/CSIC \\
Universidad Aut\'{o}noma de Madrid,
Cantoblanco, E-28049 Madrid, Spain}
}

\author{J. Portol\'es}{
  address={IFIC, Universitat de Val\`encia - CSIC, Apt. Correus 22085, E-46071 Val\`encia, Spain}
}

\author{A. Rodr\'{\i }guez-S\'anchez}{
  address={Departamento de F\'{\i }sica Te\'{o}rica 
and Instituto de F\'{\i }sica Te\'{o}rica, IFT-UAM/CSIC \\
Universidad Aut\'{o}noma de Madrid,
Cantoblanco, E-28049 Madrid, Spain}
}


\begin{abstract}
Here we study and compare the sensitivity to the Higgs sector of SUSY-seesaw
models via the LFV tau decays: $\tau \to 3 \mu$, $\tau \to \mu K^+K^-$, $\tau \to
\mu \eta$ and $\tau \to \mu f_0$. We emphasize that, at present, the two
latter channels are the most efficient ones to test 
indirectly the Higgs particles.\footnote{Talk given at the SUSY09 conference, Boston, by M. Herrero.}.
\end{abstract}

\maketitle


\section{Introduction}
Lepton Flavor Violating (LFV) tau decays provide one of the most 
efficient indirect tests of supersymmetric (SUSY) models with extended neutrino
sector, if the seesaw
mechanism for neutrino mass generation is implemented. Here we assume
SUSY-seesaw models with the MSSM particle content plus three right handed
neutrinos, $\nu_{R_i}$ $(i=1,2,3)$, 
and their
corresponding SUSY partners, ${\tilde \nu}_{R_i}$, and use 
the parameterisation for the Yukawa couplings given by 
$m_D =\,Y_\nu\,v_2 =\,\sqrt {m_N^{\rm diag}} R \sqrt {m_\nu^{\rm diag}}U^{\dagger}_{\rm
MNS}$, with 
$R$ defined by three complex angles $\theta_i$;  $v_{1(2)}= \,v\,\cos (\sin) \beta$, $v=174$ GeV; 
$m_{\nu}^\mathrm{diag}=\, \mathrm{diag}\,(m_{\nu_1},m_{\nu_2},m_{\nu_3})$ denotes the
three light neutrino masses, and  
$m_N^\mathrm{diag}\,=\, \mathrm{diag}\,(m_{N_1},m_{N_2},m_{N_3})$ the three heavy
ones. 
With this 
parameterisation it is easy to accommodate
the $\nu$ data and also get large
$Y_\nu \sim \mathcal{O}(1)$, by
choosing large $m^{\rm diag}_N$ and/or $\theta_i$.  

The sensitivity to the Higgs sector of these SUSY-seesaw models can appear only
via the LFV processes that are mediated by Higgs particles. This is the case of 
the tau decay channels considered here, whose present experimental bounds are
respectively at BR$(\tau \to 3 \mu)<3.2 \times 10^{-8}$, 
BR$(\tau \to \mu K^+K^-)<3.4 \times 10^{-8}$, BR$(\tau \to \mu \eta)<5.1 \times
10^{-8}$ and BR$(\tau \to \mu f_0)< 3.4 \times 10^{-8}$ (assuming 
BR$(f_0 \to \pi^+ \pi^-) \simeq 1$). The interest of these
channels 
is that for scenarios with heavy
SUSY soft masses of the order of 1 TeV, where the predicted rates for the
 $\tau \to \mu \gamma$ channel lay below the present 
experimental bound, BR$(\tau \to \mu \gamma) < 1.6 \times 10^{-8}$,  
still some of the Higgs-mediated processes can indeed be at the present 
experimental reach if the relevant Higgs mass is light enough, say of the order
of 100-250 GeV. We will focus here in the type of constrained SUSY-seesaw 
scenarios
called NUHM-seesaw (standing for Non Universal Higgs Mass) where this kind of
spectrum with light Higgs and heavy SUSY particles is possible. The input
parameters are $M_0$, $M_{1/2}$, $A_0$ $\tan \beta$, sign($\mu$),
$M_{H_1}=M_0(1+\delta_1)^{1/2}$ and
$M_{H_2}=M_0(1+\delta_2)^{1/2}$. In refs.~\cite{Arganda:2008jj} 
and ~\cite{Herrero:2009tm}  the proper choices of $\delta_1$ and
$\delta_2$ leading to the wanted light Higgs sector can be found. 
Notice that $\delta_1= \delta_2=0$ corresponds
to the usual constrained model (CMSSM-seesaw) with all scalar masses being
universal, but this model does not lead to the scenario with heavy SUSY and
light Higgs particles that we are interested here, so we will not considered it
next.
Most of the results 
reported here are extracted from the works
~\cite{Arganda:2008jj} and ~\cite{Herrero:2009tm} to which we
refer the reader for more details.

\section{Results and discussion}
\begin{figure}[h!]
     \begin{tabular}{cc} \hspace*{-12mm}
  	\psfig{file=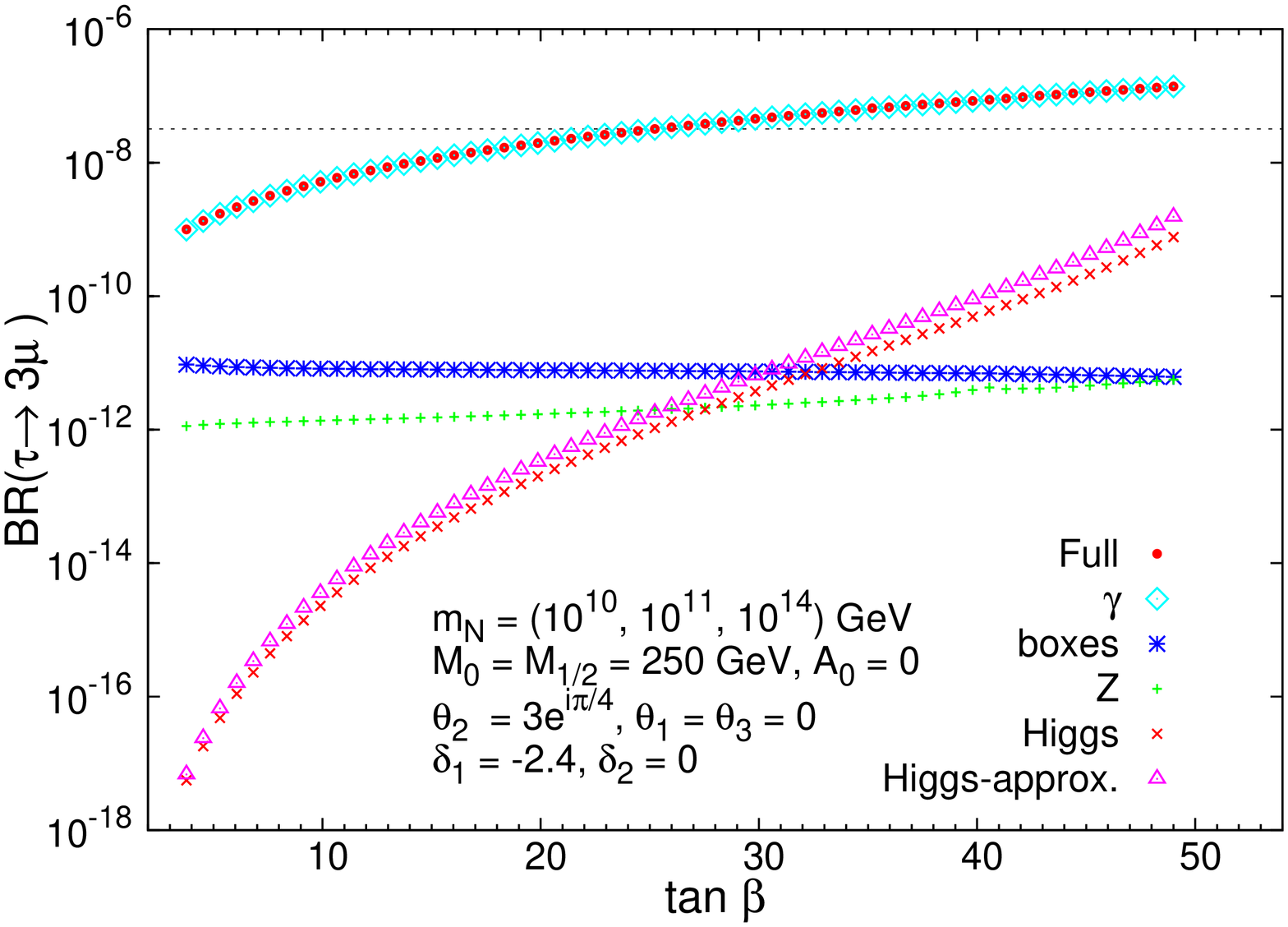,width=50mm,angle=270,clip=} 
	&
        \psfig{file=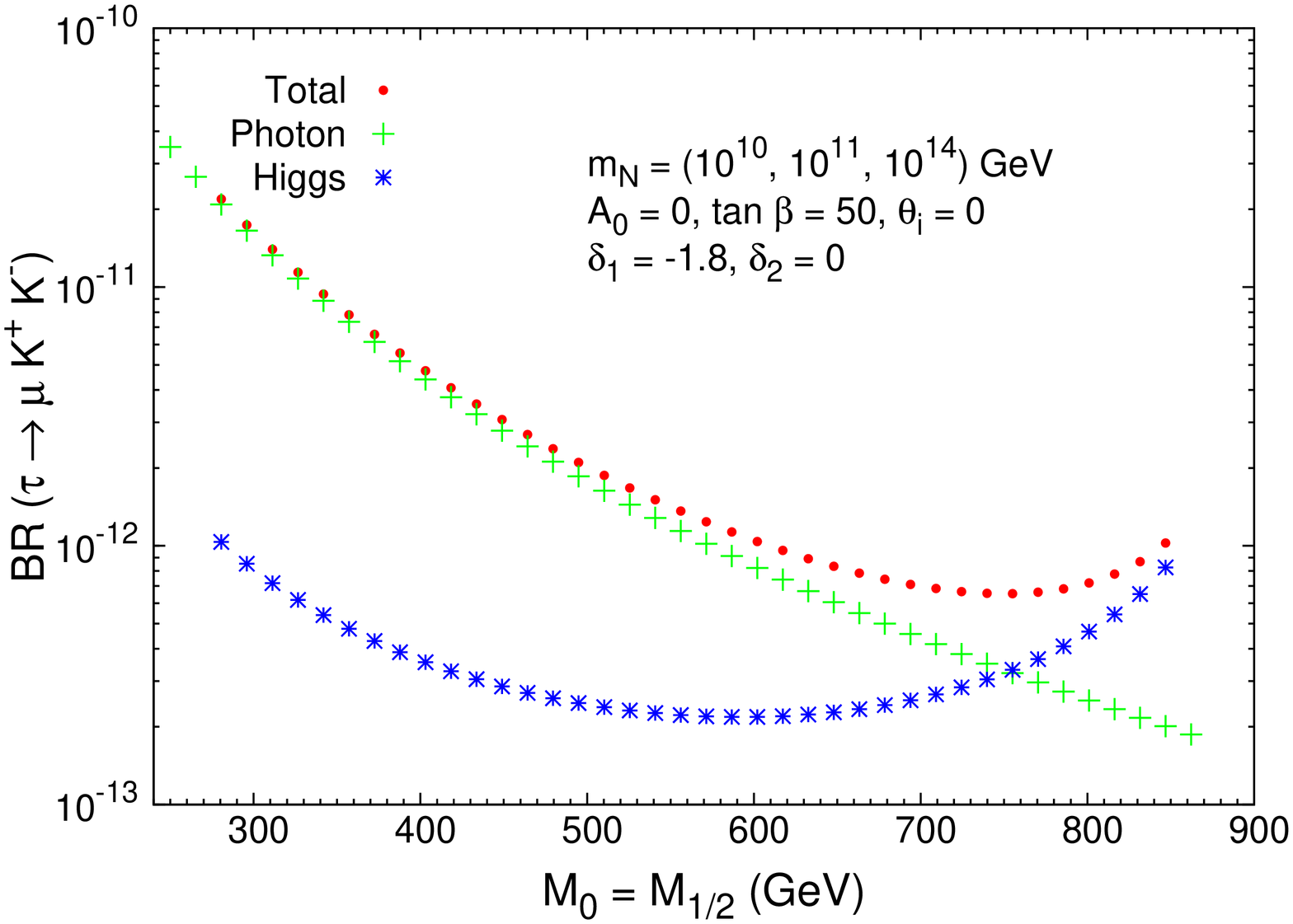,width=50mm,angle=270,clip=} \\
	\hspace*{-12mm}
	\psfig{file=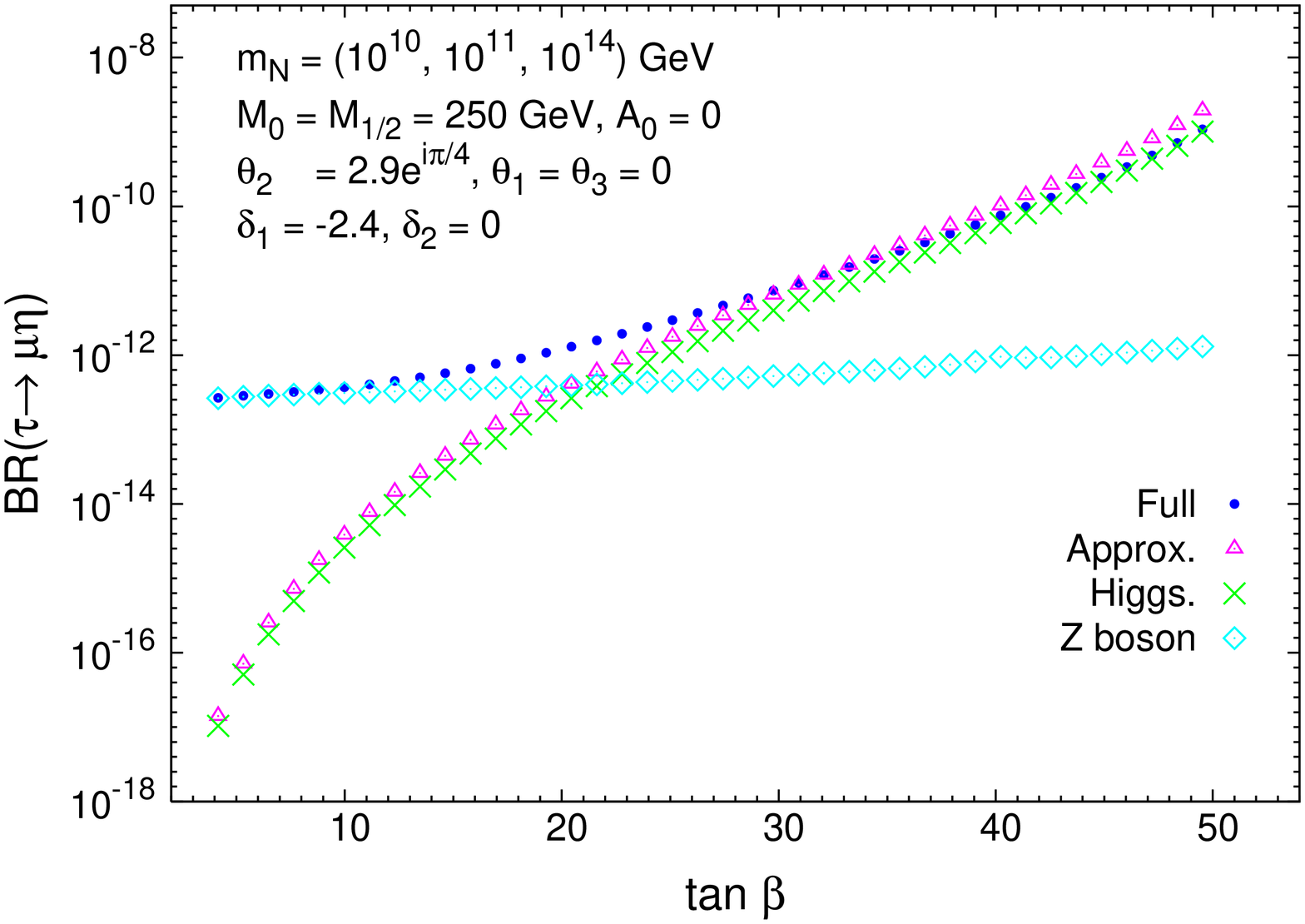,width=50mm,angle=270,clip=} 
	&
	\psfig{file=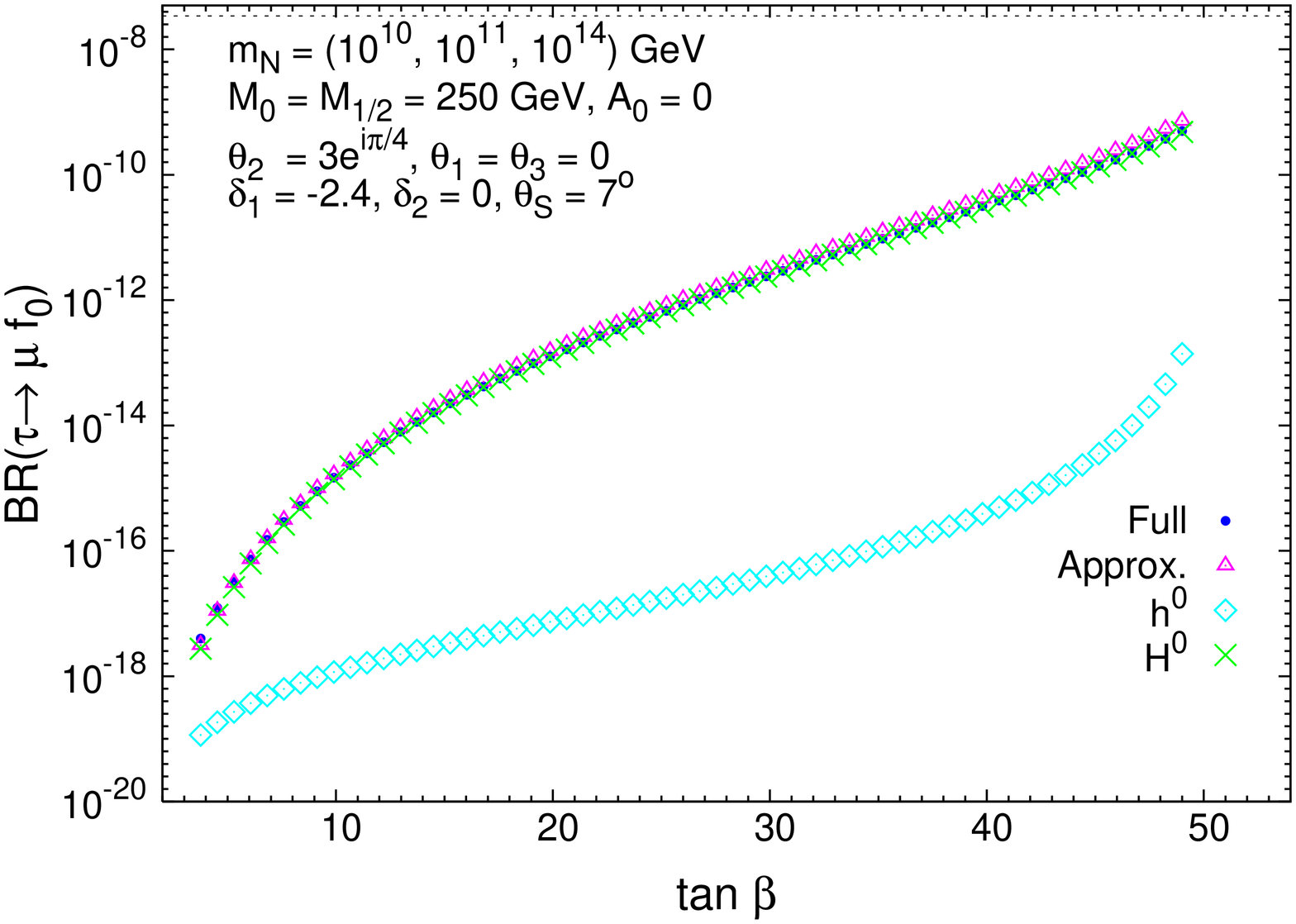,width=50mm,angle=270,clip=}	  
       \end{tabular}
     \caption{ The various contributions to 
     BR($\tau \to 3 \mu $) (upper left), BR($\tau \to \mu K^+ K^-$) 
     (upper right), BR($\tau \to \mu \eta$) (lower left) and BR($\tau \to \mu f_0$) 
     (lower right). The horizontal lines are the experimental bounds} 
     \label{fig:allBR} 
\end{figure}

The numerical results for the branching ratios of the studied LFV tau decays are
summarized in fig.~\ref{fig:allBR}. They are full one loop results 
and do not make use of any approximation like the mass insertion, large $\tan
\beta$,  nor the
leading logarithmic approximations. The mass spectra for all the involved 
particles in the loops that contribute to these processes are 
computed within the NUHM-seesaw
model, by solving the RGEs also to one loop level.  
In the case of the semileptonic decays we
have used the standard techniques in chiral theory to describe the final hadrons
in terms of quark bilinears. In particular, the channels with pseudo Goldstone
bosons (PGB), $P$, like $\pi$, $K$ and $\eta$, are treated within Chiral Perturbation
Theory ($\chi$PT) to leading order, ${\cal O}(p^2)$, where the results are given in terms
of $F_\pi= 92.4$ MeV and $m_P$. The additional contributions 
from resonances, $R$,  in
channels of the type $\tau \to \mu PP$  are taken into account within
Resonance Chiral Theory (R$\chi$T), where the results are given in terms of  
$F_\pi$, $m_P$, $m_R$
and well established form factors. In particular for the $\tau \to \mu K^+K^-$
channel the contributions from the $\rho(770)$, $\omega(782)$ and $\phi(1020)$
are considered via the electromagnetic vector form factor, $F_V^{K^+K^-}$. 
On the other hand, the $\eta(548)$ is defined via mixing between the octet, 
$\eta_8$, and singlet, $\eta_0$, components of the $P(0^-)$ nonet of PGB in 
$\chi$PT. Concretely we assume here a mixing angle of $\theta = -18^o$. The
$f_0(980)$ is defined via mixing between the octet, $R_8$, and singlet, $R_0$,
components of the $R(0^+)$ nonet of resonances in R$\chi$T. Concretely we assume
here two choices for this mixing angle, $\theta_S=7^o$ and $30^o$. Notice that
we have selected the semileptonic channels where the final hadrons have a
relevant strange quark content, and consequently the 
sensitivity to the Higgs particles is greater than in those with just up and/or
down quarks.  

\begin{figure}[h!]
     \begin{tabular}{cc} \hspace*{-12mm}
  	\psfig{file=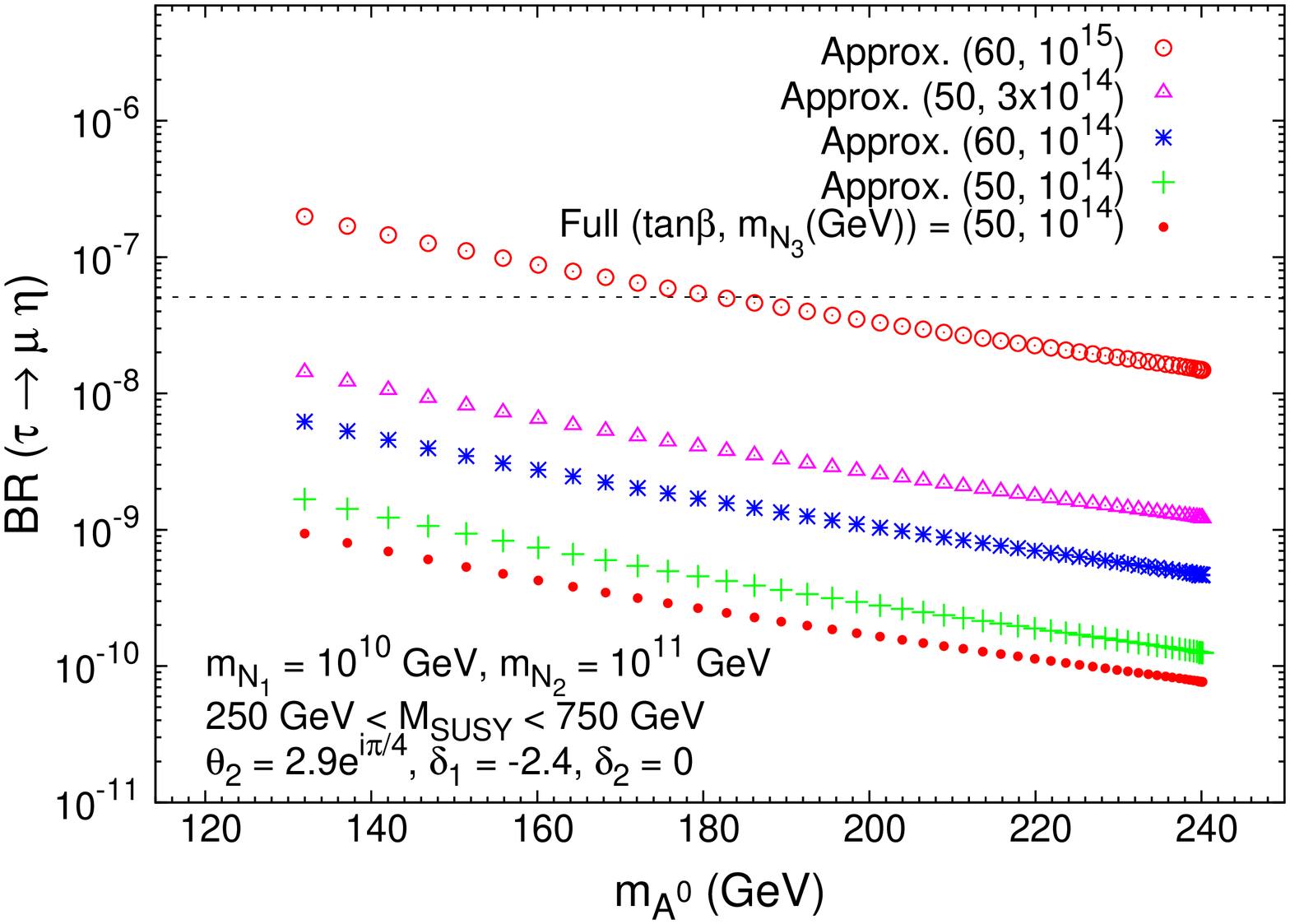,width=50mm,angle=270,clip=} 
	&
        \psfig{file=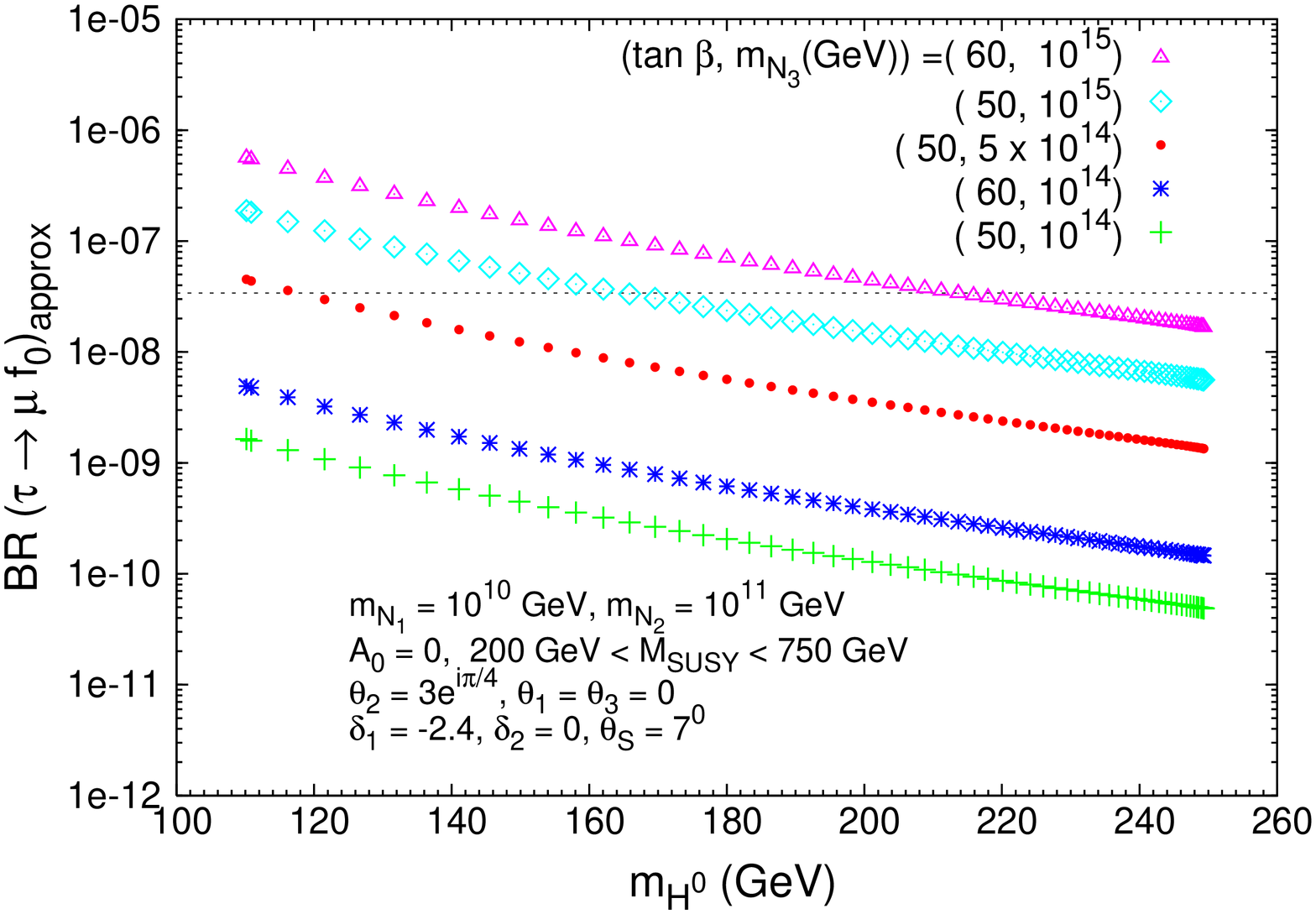,width=50mm,angle=270,clip=} 	  
       \end{tabular}
     \caption{ BR($\tau \to \mu \eta$) (left) and BR($\tau \to \mu f_0(980)$) 
        (right) 
      as a function of the relevant Higgs mass. The horizontal dashed line in
      each plot is
      the present experimental upper bound} 
     \label{fig:BReta-BRf0-mH} 
\end{figure}

Besides the total rates, we also show separately in fig.~\ref{fig:allBR} the various
contributions to these processes: 1) $\tau \to 3 \mu$ can be mediated by a
$\gamma$, a $Z$ boson, boxes and $h^0$, $H^0$ and $A^0$~\cite{Arganda:2005ji}, 
2) $\tau \to \mu K^+K^-$ by a $\gamma$ (also $Z$, but it is negligible) 
and $h^0$, $H^0$, 3) $\tau \to \mu \eta$ by a Z boson and $A^0$, and 4) 
$\tau \to \mu f_0$ by $h^0$ and $H^0$.  We conclude that although the Higgs
contributions in all these processes grow very fast with $\tan \beta$, still at
large $\tan \beta$ values
these are fairly dominated by the $\gamma$ contribution in the cases of $\tau \to
3 \mu$ and $\tau \to \mu  K^+K^-$. Therefore, these are not sensitive to the Higgs
sector. We have checked that this is true even for a very heavy SUSY spectra
where the $\gamma$ contribution gets reduced considerably. In 
$\tau \to \mu  K^+K^-$ the Higgs contribution is relevant for $M_{\rm SUSY}>750$
GeV, but there the rates are too small compared to the present bound. In contrast, the
$\tau \to \mu \eta$ and $\tau \to \mu f_0$ channels are clearly sensitive 
to the
Higgs sector. In fig.~\ref{fig:allBR} we see that the $A^0$ contribution
dominates BR($\tau \to \mu \eta$) for $\tan \beta >20$ and the $H^0$ contribution
dominates  BR($\tau \to \mu f_0$) at all $\tan \beta$ values. We also
conclude from this figure that the approximate formulas found in 
refs.~\cite{Arganda:2008jj} and ~\cite{Herrero:2009tm} for
large $\tan \beta$, whose simplest forms are given by,
\begin{eqnarray}
{\rm BR}(\tau \to \mu \eta(548))_{\rm approx}  
 &=& 1.2 \times 10^{-7} \left| \delta_{32} \right|^2 \left( \frac{100}
{m_{A^0}({\rm GeV})}
\right)^4 \left( \frac{\tan \beta}{60} \right)^6 \nonumber \\[2mm]
 {\rm BR}(\tau \to \mu f_0(980))_{{\rm approx}}  
 &=& \! \! \! \left( \!  \begin{array}{c} 7.3 \times 10^{-8}\,\, (\theta_S=7^\circ) 
\\ 
4.2 \times 10^{-9}\,\, (\theta_S=30^\circ)  \end{array} \! \! \right) 
\left| \delta_{32} \right|^2    \left( \frac{100}
{m_{H^0}({\rm GeV})}
\right)^4 \left( \frac{\tan \beta}{60} \right)^6  \! \! , 
 \nonumber 
\end{eqnarray}
 provide a very good approximation to the full result. This is also shown in 
 fig.\ref{fig:BReta-BRf0-mH} where BR($\tau \to \mu \eta$) and 
 BR($\tau \to \mu f_0(980)$) are displayed as a function of the relevant Higgs
 mass. We see clearly in this figure that these two channels are  
 sensitive to masses within the range 100-250 GeV,  
 for large $\tan \beta$, $\theta_2$ and $m_{N_3}$.     
\begin{figure}[h!]
     \begin{tabular}{ccc} \hspace*{-12mm}
  	\psfig{file=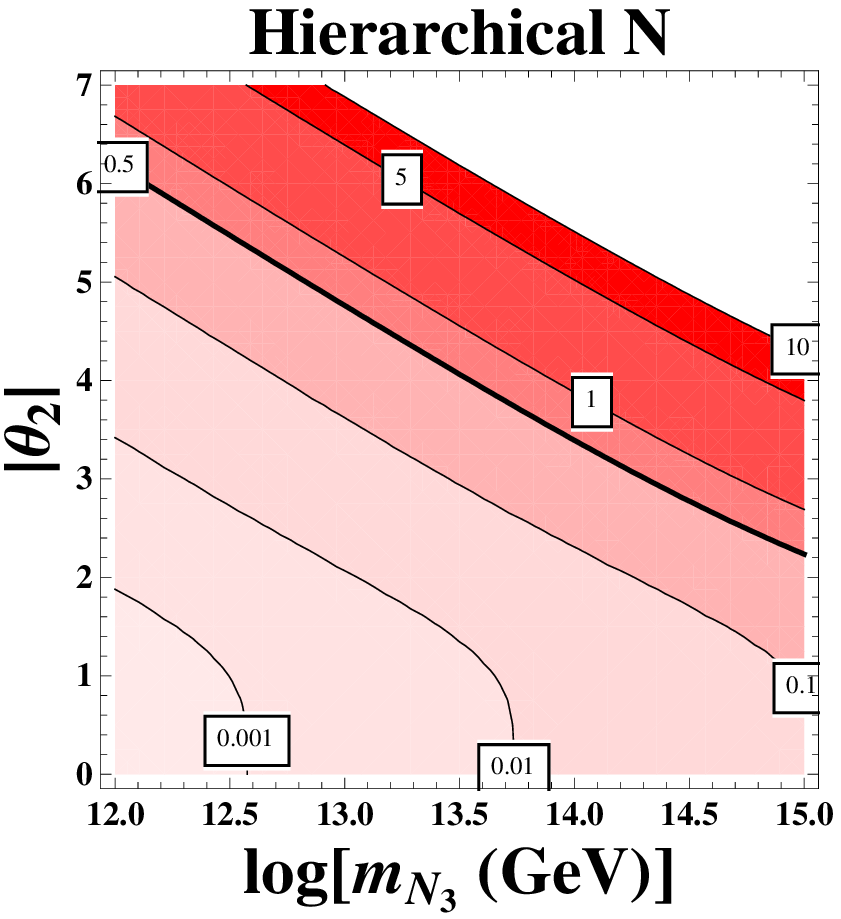,width=50mm,clip=} 
	&
	\psfig{file=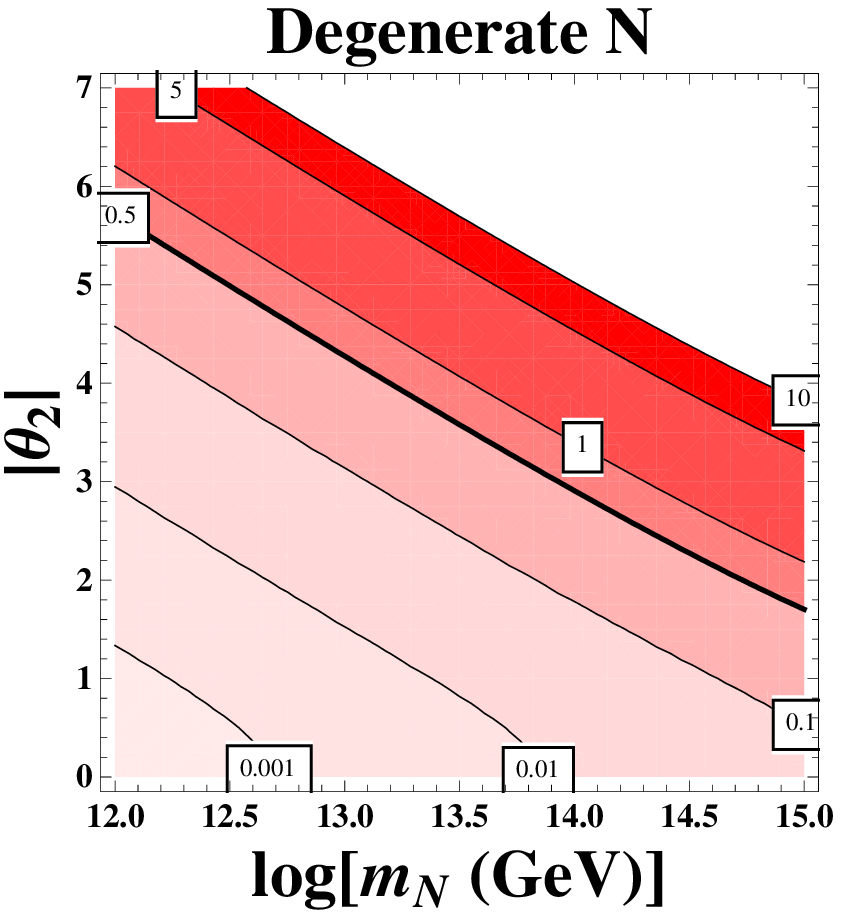,width=50mm,clip=}
	&
        \psfig{file=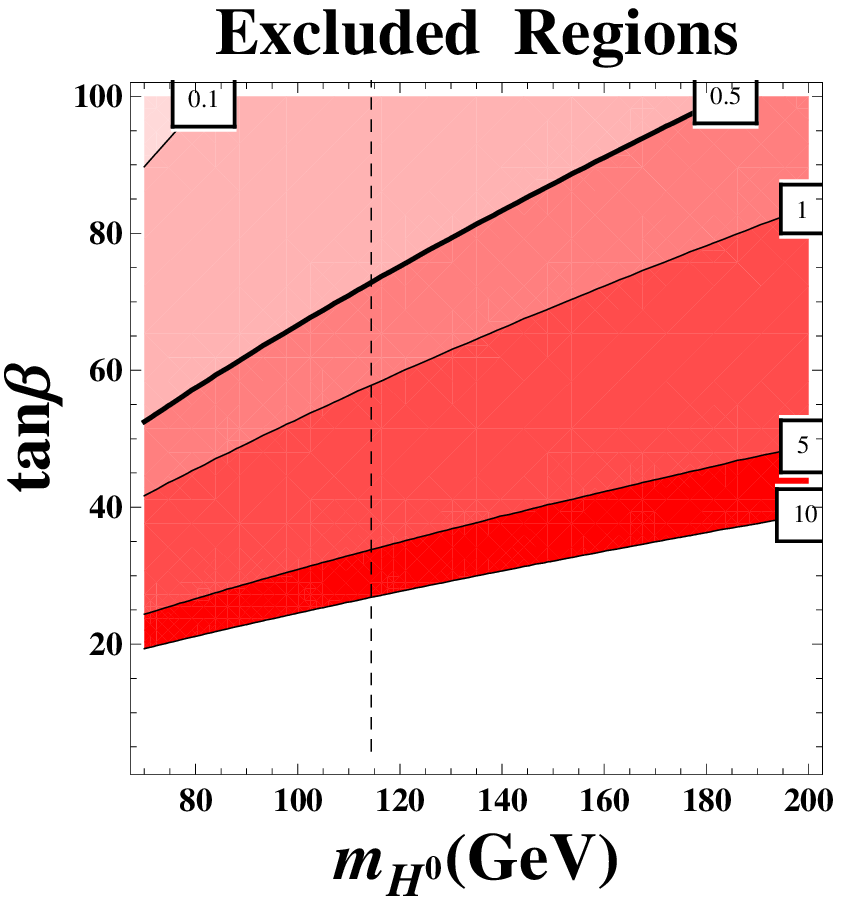,width=52mm,clip=} 	  
       \end{tabular}
     \caption{Left (central) panel: contours of $|\delta_{32}|$ in SUSY-seesaw for
     hierarchical (degenerate) heavy neutrinos. Right panel: Excluded regions in the 
     ($m_{H^0}$, $\tan \beta$) plane from the study of $\tau \to \mu f_0$.
     The exluded areas are 
       those above the 
     contour lines corresponding to fixed $|\delta_{32}|=0.1,0.5,1,5,10$.} 
     \label{fig:final} 
\end{figure}
Finally, fig.\ref{fig:final} illustrates several examples for the relevant 
parameter $\delta_{32}$ that meassures approximately the size of the LFV in the tau-mu sector 
in seesaw scenarios with a) hierarchical and b) degenerate heavy neutrinos $N$.
We see that in both scenarios values as large as $|\delta_{32}|~\sim 1-10$ 
can be obtained. Therefore, with such large values and the present 
experimental upper limits one can extract lower bounds for the relevant 
Higgs mass and upper bounds for $\tan \beta$. This is the main conclusion of
this work. The case of $\tau \to \mu f_0$ is illustrated in the last plot 
of fig.\ref{fig:final}, where one can see the excluded regions in the 
$(m_{H^0}, \tan \beta)$ plane. 
\begin{theacknowledgments}
M. Herrero acknowledges the SUSY09 organisers for her
invitation to give this talk and for the fruitful conference.   
\end{theacknowledgments}


\begin{thebibliography}{9}
\bibitem{Arganda:2008jj}
  E.~Arganda, M.~J.~Herrero and J.~Portoles,
  JHEP {\bf 0806} (2008) 079
  [arXiv:0803.2039 [hep-ph]].
\bibitem{Herrero:2009tm}
  M.~J.~Herrero, J.~Portoles and A.~M.~Rodriguez-Sanchez,
  Phys.\ Rev.\  D {\bf 80}, 015023 (2009)
  [arXiv:0903.5151 [hep-ph]].
\bibitem{Arganda:2005ji}
E.~Arganda and M.~J.~Herrero,
Phys.\ Rev.\  D {\bf 73} (2006) 055003
[arXiv:hep-ph/0510405].
\end{thebibliography}
\end{document}